\documentclass [12pt]{article}

\usepackage {graphicx}

\begin{document}
Reported at the NATO Advanced Study Institute on Physics and Materials Science of Vortex States, Flux Pinning and Dynamics, Kusadasi, Turkey, July 26 - August 8, 1998, published by Kluwer Academic Publishers, R. Kossowski at al., eds.  p.609-629, 1999 

\bigskip

\textbf{WHAT IS THE VORTEX LATTICE MELTING, REALITY OR FICTION?}

\bigskip

A.V.NIKULOV

\textit{Institute of Microelectronics Technology and High Purity Materials, Russian Academy of Sciences, 142432 Chernogolovka, Moscow Region, Russia.}

\bigskip

\begin{center}
\textbf{Abstract}
\end{center}
The concept of the transition into the Abrikosov state as the appearance of  long-range phase coherence is considered. It is shown that the famous Abrikosov solution gives qualitatively incorrect result. The transition into the Abrikosov state should be first order in ideal (without disorder) superconductor. Such a transition is observed only in bulk superconductors with weak disorders. The absence of a sharp transition in thin films with weak disorder is interpreted as the absence of long-range phase coherence down to a very low magnetic field. The observed smooth phase coherence appearance in superconductors with strong disorder is explained by the increase of the effective dimensionality. It is claimed that no experimental evidence of the vortex lattice melting exists now. It is shown that this popular concept has appeared on a consequence of incorrect conception about the Abrikosov state and incorrect definition of the phase coherence.

\bigskip

\textbf{1. Introduction}

\bigskip

The concept of vortex lattice melting has become very popular after the discovery of high-Tc superconductors (HTSC). Hundreds of papers have appeared in the last ten years in which the vortex lattice melting is considered. Some papers with titles "Evidence of the vortex lattice melting ...." were published in last years. Different features in different superconductors were interpreted as the vortex lattice melting. But the reality of the vortex lattice melting is not evident at present. 

A feature can be interpreted as a universal transition if it is observed in all superconductors. The vortex lattice melting is a universal transition. No peculiarity of HTSC or low-Tc (conventional) superconductors (LTSC) is considered in the theories of the vortex lattice melting [1]. Therefore, if this transition exist it must be observed both in HTSC and in LTSC. The position difference of the transition in HTSC and in LTSC should correspond to the scaling law [2]. Three-dimensional (bulk) and two-dimensional 
superconductors ought to be distinguish at the comparison of the investigation results of HTSC and LTSC. Layered superconductors (both HTSC 
and LTSC) ought to be considered as an intermediate case between three- and 
two-dimensional superconductors. 

The comparison of the investigation results of HTSC and LTSC shows that the 
sharp change of the resistive properties observed in bulk superconductors 
with weak disorder only can be interpreted as an universal transition. This 
feature is observed both in conventional superconductors [3] and in 
YBa$_{{\rm 2}}$Cu$_{{\rm 3}}$O$_{{\rm 7}{\rm -} {\rm x}}$ [4]. The position difference of this transition in these superconductors conforms to the scaling law [2,5]. I have denoted [5] the position of this transition as H$_{{\rm c}{\rm 4}}$.

This transition was interpreted in the paper [4] as the vortex lattice melting, but in the work [3], in which it was observed first, it was interpreted as a transition from the mixed state without phase coherence (called as "one-dimensional" state in this paper) into the Abrikosov state, considered as the mixed state with long-range phase coherence. It is contended in the present paper that the interpretation [3] is right and the interpretation [4] can not be right. 

The concept of vortex lattice melting has become popular in consequence of some no enough correct conception about the Abrikosov state. Therefore this knowledge is defined more exactly in the Section 2. The knowledge about the mixed state of most scientists was formed by the influence of results of the mean field approximation. Results obtained in the mean field approximation 
and in the fluctuation theory are compared in the Section 3. It is shown 
that according to the fluctuation theory the transition into the Abrikosov 
state of an ideal superconductor must be first order. Results of 
experimental investigation of the transition into the Abrikosov state in 
bulk superconductors with weak disorders, in thin films (two-dimensional 
superconductors) and in superconductors with strong disorder are considered 
in the Section 4. Theories of the vortex lattice melting and of the vortex 
liquid solidification are reviewed briefly in the Section 5.

\

\bigskip

\textbf{2. The Abrikosov State is the Mixed State with Long-Range Phase Coherence.} 

\bigskip

According to a popular point of view [6] the pinning effect can be in the 
vortex lattice only. The process of the vortex pinning is very different for 
vortex liquids or solids. In the case of a vortex solid, a few pinning 
centers can hold the entire lattice because it is stiff. But it is 
impossible to hold a vortex liquid in place with the help of a few pinning 
centers. Under the influence of this point of view the absence of the vortex 
pinning in wide region below H$_{{\rm c}{\rm 2}}$ in HTSC was interpreted by 
many researchers as consequence of the vortex lattice melting.

The point of view [6] became popular in a consequence of the wide spread 
opinion that the Abrikosov state is the flux line lattice (FLL) [1] like an 
atom lattice, or a lattice of long molecules [7] and the Abrikosov vortex is 
magnetic flux. The transport properties were connected with the motion of 
the magnetic flux structures (or FLL) induced by the Lorentz force [8].

Therefore it is important to emphasize that the Abrikosov state is the mixed 
state with long-range phase coherence, vortices are singularities in this 
state but no magnetic flux and that the resistivity in the Abrikosov state 
is caused by the vortex flow but not by the magnetic flux flow. 

\bigskip

2.1.\textbf{} THE ABRIKOSOV VORTICES ARE SINGULARITIES IN THE MIXED STATE WITH LONG-RANGE PHASE COHERENCE

\bigskip

Superconductivity is a macroscopic quantum effect. The phase $\varphi $ of the wave function $\Psi $= $\vert \Psi \vert $exp(i$\varphi $) of superconducting electrons can cohere all over the volume of a superconductor. According to the relation for the velocity of the superconducting electrons (see for example [8]) the relation for the superconducting current j$_{{\rm s}}$ = n$_{{\rm s}}$ev$_{{\rm s}}$

\begin{equation}
\label{eq1}
{\oint\limits_{l} {dl\lambda _{L}^{2} j_{s} = {\frac{{\Phi _{0}} }{{2\pi 
}}}{\oint\limits_{l} {dl\nabla \varphi - \Phi} } }} 
\end{equation}

\noindent
is valid in the region where the phase coherence exists. Here A is the vector potential; $\Phi _{{\rm 0}}$ = hc/2e is the flux quantum; n$_{{\rm 
s}}$ is the superconducting electron density; $\lambda _{{\rm L}}$ = 
(mc/e$^{{\rm 2}}$n$_{{\rm s}}$)$^{{\rm 0}{\rm .}{\rm 5}}$ is the London 
penetration depth; l is a closed path of integration; $\Phi = 
{\oint\limits_{l} {dlA}} $ is the magnetic flux contained within the closed 
path of integration l. 

If a singularity is absent, then ${\oint\limits_{l} {dl\nabla \varphi = 0} 
}$. In this case the relation (\ref{eq1}) is the equation postulated by F. and H.London for the explanation of the Meissner effect (see [8]). The magnetic 
flux can not penetrate in this case inside a superconductor with long-rang phase coherence. 

Consequently, a magnetic field can penetrate into a superconductor only if: 1) superconductivity is destroyed, or 2) singularities appears, or 3) the 
phase coherence is absent. The first is observed in type I superconductors. 
The second case is observed in the Abrikosov state of type II 
superconductors. The singularities are the Abrikosov vortices in this case. 
${\frac{{1}}{{2\pi} }}{\oint\limits_{l} {dl\nabla \varphi = n}} $ is the 
number of the Abrikosov vortices contained within the closed path of 
integration; n = $\Phi $/$\Phi _{{\rm 0}}$ if l is large. These two cases 
are widely known. The third case is not so well known. It is observed in fluctuation superconductivity region, for example above the second critical field, H$_{{\rm c}{\rm 2}}$, in type II superconductors (see below).

Consequently, the Abrikosov vortices are no flux lines but are singularities in the mixed state with phase coherence. These singularities allow the magnetic flux to penetrate into a superconductor. A singularity can not exist without a medium. Consequently the existence of the vortices is an evidence of the existence of the phase coherence.

\

2.2. THE VORTEX FLOW RESISTIVITY

\bigskip

According to the Josephson relation (see for example [9]), a electric voltage in the Abrikosov state is caused by the vortex flow. The phase difference $\varphi $ between two points changes by 2$\pi $ when it crosses a line connecting these points. Consequently, the macroscopic voltage in the Abrikosov state is equal to $E = (v_{{\rm v}{\rm o}{\rm r}} \times  n_{{\rm v}{\rm o}{\rm r}}\Phi _{{\rm 0}})/c$. Where $v_{{\rm v}{\rm o}{\rm r}}$ is the velocity of the vortex flow, $n_{{\rm v}{\rm o}{\rm r}}$ is the density of the vortices.

This result coincides nominally with the one obtained by the Faraday's law $E = -( v_{{\rm v}{\rm o}{\rm r}} \times  B)/c$, because $B = n_{{\rm v}{\rm o}{\rm r}}\Phi _{{\rm 0}}$ in the Abrikosov state. Therefore, it became possible that the resistivity in the Abrikosov state is considered as a consequence of flux flow in all textbooks (see for example [8,9]) and majority of papers, although it is obvious that the magnetic flux does not flow in a superconductors in this case. It is obvious also that the Lorentz force can not be the driving force on an Abrikosov vortex [10]. The notation "flux flow resistance" is not quite correct. I will use the more correct notation "vortex flow resistivity" instead of "flux flow resistivity".

\bigskip

2.3. THE VORTEX PINNING

\bigskip

The resistivity can be equal zero in the Abrikosov state if the pinning force prevents from the vortex flow [8]. The vortex pinning is a consequence of superconductor disorders. Because the vortexes are singularities in the mixed state with the long-rang phase coherence, the vortex pinning is a consequence of the phase coherence also. The vortex pinning can not exist in a state without the phase coherence. 

Thus, the Abrikosov state is the mixed state with long-range phase coherence. Changes of the resistive properties should be observed first of all at the transition into the Abrikosov state, because a transition from the paraconductivity regime to the vortex flow regime must occur and the vortex pinning can appear at the appearance of the long-range phase coherence.

\bigskip

\textbf{3. Transition into the Abrikosov State in the Mean Field Approximation and in the Fluctuation Theory.}

\bigskip

Thermodynamic average values are calculated in the fluctuation theory. The thermodynamic average of any quality, for example superconducting electron density 

\begin{equation}
\label{eq2}
 < \vert \Psi \vert ^{2} > = {\frac{{\sum {\vert \Psi \vert ^{2}\exp ( - F_{GL} / k_{B} T)}} }{{\sum {\exp ( - F_{GL} / k_{B} T)}} }}
\end{equation}

\noindent
can be calculated in different approximation in different real cases. We can write the relation of the Ginzburg-Landau free energy to $k_{{\rm B}}T$ in a dimensionless unit system [2]

\begin{equation}
\label{eq3}
{\frac{{F_{GL}} }{{k_{B} T}}} = {\sum\limits_{k} {(\varepsilon _{n} + q_{z}^{2} )\vert \Psi _{k} \vert ^{2} + {\frac{{1}}{{2V}}}{\sum\limits_{k_{i}}  {V_{k_{1} k_{2} k_{3} k_{4}}  \Psi _{k_{1}} ^{\ast}  \Psi _{k_{2}} ^{\ast}  \Psi _{k_{3}}  \Psi _{k_{4}} } } } }
\end{equation}
\noindent

Here the wave function $\Psi $(r) of superconducting electrons is expanded by the eigen functions $\varphi _{{\rm k}}(r)$: $\Psi (r) = V^{ - 1 / 2}{\sum\limits_{k} {\Psi _{k} \varphi _{k} (r)}} $; $k = (n,l,q_{{\rm z}})$; $n$ is the number of the Landau level; $l$ is the index of the Landau level functions; $q_{{\rm z}}$ is the longitudinal (along magnetic field) wavevector; $V$ is the superconductor volume; $\varepsilon _{{\rm n}} = (t-1+h+2nh)/Gi_{{\rm H}}= (h-h_{{\rm c}{\rm 2}}+2nh)/Gi_{{\rm H}}$; $Gi_{{\rm H}} = Gi^{{\rm 1}{\rm /} {\rm 3}}(th)^{{\rm 2}{\rm /} {\rm 3}}$ is the Ginzburg number in magnetic field; $Gi = (k_{{\rm B}}T_{{\rm c}}/H_{{\rm c}}(0)\xi ^{{\rm 3}}(0))^{{\rm 2}}$ is the Ginzburg number; $t = T/T_{{\rm c}}$; $h = H/H_{{\rm c}{\rm 2}}(0)$; $V_{k_{1} k_{2} k_{3} k_{4}}  = V^{ - 1}{\int\limits_{V} {d^{3}r\varphi _{k_{1}} ^{\ast}  \varphi _{k_{2}} ^{\ast}  \varphi _{k_{3}}  \varphi _{k_{4}} } } $.

\

3.1. THE ABRIKOSOV SOLUTION

\bigskip

States corresponded to minimum value of the free-energy (3) are considered only in the mean field approximation. At $H_{{\rm c}{\rm 2}}/3 < H < H_{{\rm c}{\rm 2}}$ a contribution by the lowest (n=0) Landau level (LLL) only can decrease the free energy value because $\varepsilon _{{\rm n}{\rm =} {\rm 0}} < 0$ and $\varepsilon _{{\rm n}{\rm =} {\rm 1}} > 0$ in this region. The states with $q_{{\rm z}} = 0$ correspond to the minimum value of the free energy. Therefore the lowest Landau level and zero-dimensional case is considered in the mean field approximation. The Ginzburg-Landau free energy can be written as 

\begin{equation}
\label{eq4}
{\frac{{F_{GL}} }{{k_{B} T}}} = V(\varepsilon \overline {\vert \Psi \vert ^{2}} + {\frac{{\beta _{a}} }{{2}}}\overline {\vert \Psi \vert ^{2}} ^{2})
\end{equation}

\noindent
in this case. Here $\varepsilon  = \varepsilon _{{\rm n}{\rm =} {\rm 0}} = (h-h_{{\rm c}{\rm 2}})/Gi_{{\rm H}}$; $\overline {\vert \Psi \vert ^{2}} = V^{ - 1}{\int\limits_{V} {d^{3}r\vert \Psi (r)\vert ^{2}}} $ is the spatial average density of superconducting electrons; $\beta _{a} = \overline {\vert \Psi \vert ^{4}} / \overline {\vert \Psi \vert ^{2}} ^{2}$ is the Abrikosov parameter; $\overline {\vert \Psi \vert ^{4}} = V^{ - 1}{\int\limits_{V} {d^{3}r\vert \Psi (r)\vert ^{4}}} $. The values $\beta _{{\rm a}} = \beta _{{\rm A}}$ and $\vert \Psi \vert ^{{\rm 2}} = \varepsilon /\beta _{{\rm A}}$ correspond to the minimum of the 
Ginzburg-Landau free-energy, where $\beta _{{\rm A}}$ is a minimum possible value of $\beta _{{\rm a}}$. The minimum value $\beta _{{\rm A}}  \approx  1.16$ corresponds to the triangular vortex lattice [11] if $\Psi (r)$ are the functions of the lowest Landau level. This state is known as the Abrikosov state, although he considered the square vortex lattice with $\beta _{{\rm A}}  \approx  1.18$ [12]. According to the Abrikosov solution [12] a finite density of superconducting electrons, long-rang phase coherence and the vortex lattice appear simultaneously at $H = H_{{\rm c}{\rm 2}}$. This transition was considered as second order phase transition [13].

\

3.2. THE MIXED STATE WITHOUT PHASE COHERENCE.

\bigskip

The long-rang phase coherence is not connected directly with the density of the superconducting electrons in the fluctuation theory. Therefore a mixed state without phase coherence can exist according to this theory. Because the Abrikosov state is the mixed state with long-range phase coherence the transition into the Abrikosov state is an appearance of long-range phase coherence.

This transition is not connected with H$_{{\rm c}{\rm 2}}$ where a crossover from low to high density of the superconducting electrons occurs. This means that the "jump" of the specific heat and the fracture of the magnetization dependence at $H_{{\rm c}{\rm 2}}$ are not connected with the transition into the Abrikosov state because these thermodynamic properties depend on the spatial average density of the superconducting electrons only.

\bigskip

3.3. REDACTION OF THE EFFECTIVE DIMENSIONALITY OF THE FLUCTUATION ON TWO IN A MAGNETIC FIELD 

\bigskip

The mixed state without the phase coherence differs qualitatively from the fluctuation state above $T_{{\rm c}}$ because the effective dimensionality of the fluctuation decreases on two near $H_{{\rm c}{\rm 2}}$ [14]. This means that only an longitudinal (along magnetic field direction) component, $\xi _{{\rm z}}$, of the length of the phase coherence increases near $H_{{\rm c}{\rm 2}}$. Whereas all components of the coherence length, i.e. $\xi _{{\rm x}}$, $\xi _{{\rm y}}$, $\xi _{{\rm z}}$, increase up to the infinity at $T_{{\rm c}}$ (see Fig.1). Therefore the second order phase transition takes place at $T_{{\rm c}}$ and an phase transition at $H_{{\rm c}{\rm 2}}$ is absent. The second order phase transition can not take place in the one-dimensional system. Thus, the transition into the Abrikosov state differs qualitatively from the transition into the superconducting state in zero magnetic field. 

\begin{figure}
\centering 
\includegraphics[width=3.53in,height=2.85in]{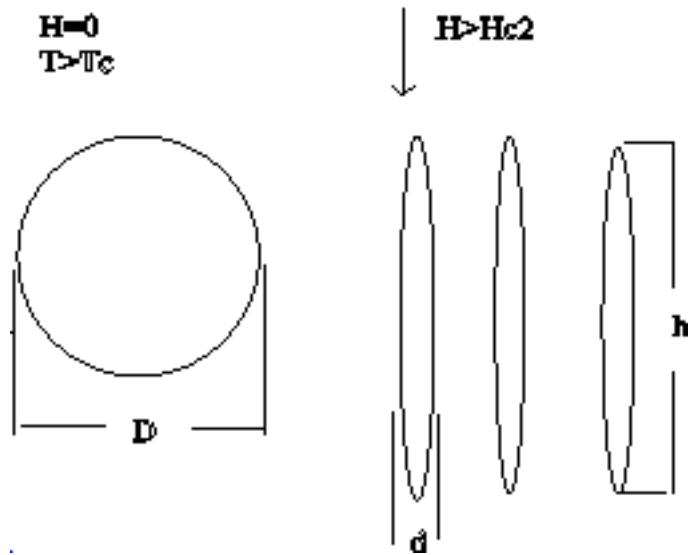}
\caption{ Superconducting drops above $T_{{\rm c}}$ and above $H_{{\rm c}{\rm 2}}$. $D = \xi (T) = \xi (0)(T/T_{{\rm c}} - 1)^{{\rm -} {\rm 1}{\rm /} {\rm 2}}$; $h = \xi _{{\rm z}} = (\Phi _{{\rm 0}}/2\pi (H-H_{{\rm c}{\rm 2}}))^{{\rm 1}{\rm /} {\rm 2}}$; $d = \xi _{{\rm R}}  \approx  (\Phi _{{\rm 0}}/H)^{{\rm 1}{\rm /} {\rm 2}}$ }
\end{figure}

\

3.4. TRANSITION INTO THE ABRIKOSOV STATE OF IDEAL SUPERCONDUCTOR MUST BE FIRST ORDER

\bigskip

The length of the phase coherence in the mixed state without long-range phase coherence can be defined from the correlation function (see [9]). In a high magnetic field (in the LLL region) in the linear approximation the longitudinal component is equal $\xi _{{\rm z}} = (\Phi _{{\rm 0}}/2\pi (H-H_{{\rm c}{\rm 2}}))^{{\rm 1}{\rm /} {\rm 2}}$ and transversal component is equal $\xi _{{\rm R}} = (2\Phi _{{\rm 0}}/\pi H)^{{\rm 1}{\rm /} {\rm 2}}$. The relation $\xi _{{\rm z}} = (\Phi _{{\rm 0}}/2\pi (H-H_{{\rm c}{\rm 2}}))^{{\rm 1}{\rm /} {\rm 2}{\rm} }$can be valid in the linear approximation only and must be renormalized near $H_{{\rm c}{\rm 2}}$. Whereas the transversal component changes little near $H_{{\rm c}{\rm 2}}$ and $\xi _{{\rm R}}$ value is close to $(\Phi _{{\rm 0}}/H)^{{\rm 1}{\rm /} {\rm 2}}$ in the whole LLL region.

Thus, if we define the phase coherence by the correlation function we can conclude that the long-rang phase coherence can not be in the mixed state of type II superconductors. On other hand we know that the Abrikosov state is the mixed state with long-rang phase coherence. Consequently, the definition of the phase coherence by the correlation function is unsuited for the mixed state. It must follow from a right definition that the existence of the singularities (the Abrikosov vortexes) is evidence of the phase coherence.

We can use the relation (\ref{eq1}) for the definition of the phase coherence: the phase coherence exists in some region if the relation (\ref{eq1}) is valid for any closed path in this region. It is obvious that according to this definition the long-rang phase coherence exists both in the Meissner and Abrikosov states.

Because in the mixed state without the phase coherence the transversal length of the phase coherence $\xi _{{\rm R}} = (\Phi _{{\rm 0}}/H)^{{\rm 1}{\rm /} {\rm 2}}$ changes little near $H_{{\rm c}{\rm 2}}$ two characteristic lengths only ($(\Phi _{{\rm 0}}/H)^{{\rm 1}{\rm / }{\rm 2}}$ and sample size $L$) are across magnetic field direction in a superconductor without disorder. Consequently, the length of the phase coherence changes by jump from $(\Phi _{{\rm 0}}/H)^{{\rm 1}{\rm /} {\rm 2}}$ to $L$ at the transition into the Abrikosov state. Thus, the result obtained in the fluctuation theory differs qualitatively from the one obtained in the mean field approximation. The transition into the Abrikosov state can not be second order phase transition. It must be first order phase transition in ideal superconductors without disorder.

\bigskip

3.5 MAKI-TAKAYAMA RESULT 

\bigskip

According to the mean field approximation [12] the Abrikosov state is the mixed state with long-range order and the vortex lattice. Consequently, if the mean field approximation is valid in the thermodynamic limit, then two long-range orders exist in the Abrikosov state: phase coherence and crystalline long-range order of the vortex lattice. The mean field approximation is valid if the fluctuation correction is small.

Eilenberger [15] has proposed a method of the calculation of the fluctuation correction to the Abrikosov solution. Using this method Maki and Takayama [16] have shown that the fluctuation correction $\Delta n _{{\rm s}{\rm ,}{\rm f}{\rm l}}$ to the Abrikosov solution calculated in the linear approximation depends on superconductor size $L$ across magnetic field direction: $\Delta n _{{\rm s}{\rm ,}{\rm f}{\rm l}}$ is proportional  $\ln(L/\xi) $ in three-dimensional superconductor and $\Delta n_{{\rm s}{\rm ,}{\rm f}{\rm l}}$ is proportiona $(L/\xi )^{{\rm 2}}$ in two-dimensional superconductor. 

The Maki-Takayama result [16] seems very queer for most scientists, because they think that it contradicts to the direct observation [17] of the vortex lattice. But the direct observation [17] of the vortex lattice is not evidence of the crystalline long-range order. If we lay along a fishing net with help of stakes it will look as a lattice. But from this direct observation we can not draw a conclusion that the fishing net is a lattice which can melt. Larkin [18] has shown that the crystalline long-rang order of the vortex lattice is unstable against the introduction of random pinning. Consequently, we can not contend on base of direct observation [17] that the Abrikosov state is the vortex lattice which can melt because it can be a structure like the fishing net.

Almost nobody has believed in a reality of the Maki-Takayama result, even author. Maki (with Thompson) attempted to correct this result in [19]. In spite of the opinion of many scientists (see [1]) I claim that the Maki-Takayama result [16] is right because it is confirmed by experimental results [20,21].

\bigskip

3.6. SCALING LAW IN LOWEST LANDAU LEVEL APPROXIMATION

\bigskip

According to (\ref{eq2}) and (3), in the lowest Landau level (LLL) approximation the thermodynamic average value of superconducting electron density, $n_{{\rm s}}$, is function of temperature, magnetic field and superconductor parameters through the $\varepsilon $ value only. Consequently, a quantity depended on the $n_{{\rm s}}$ value only, for example specific heat and magnetization, must be a universal function of $\varepsilon  = (t-1+h)/Gi_{{\rm H}}$ in the region $h \gg  Gi$, $h-h_{{\rm c}{\rm 2}} \ll  2h$, $h > h_{{\rm c}{\rm 2}}/3$, where the LLL approximation is valid. Transport properties, for example the resistivity, depend not only on the $n_{{\rm s}}$ value but also on the length of phase coherence. Therefore they can follow to the scaling law in the mixed state without phase coherence only. 

If a transition takes place at $\varepsilon _{{\rm c}{\rm 4}}$ its position $h_{{\rm c}{\rm 2}} - h_{{\rm c}{\rm 4}}$ must be a universal function of the temperature and superconductor parameters. The scaling law is general consequence of the fluctuation Ginzburg-Landau theory. Consequently, all theories obtained on base of the fluctuation Ginzburg-Landau theory must predict the same $h_{{\rm c}{\rm 2}} - h_{{\rm c}{\rm 4}}$ dependence on the temperature and superconductor parameters in 
the LLL approximation region. 

\bigskip

3.7. CHANGE OF THE RESISTIVE PROPERTIES AT THE TRANSITION INTO THE ABRIKOSOV STATE

\bigskip

According to the mean field approximation the transition to the vortex flow regime occurs at $H_{{\rm c}{\rm 2}}$. According to this approximation the vortex flow resistivity, $\rho _{{\rm f}}$, near $H_{{\rm c}{\rm 2}}$ can be described by the relation 
\begin{equation}
\label{eq5}
{\frac{{\rho _{f}} }{{\rho _{n}} }} = \gamma (1 - {\frac{{H}}{{H_{c2}} }})
\end{equation}
 $\rho _{{\rm n}}$ is the resistivity in the normal state. The coefficient $\gamma $ was calculated in many works (see [22]).

According to (\ref{eq4}) the resistivity of ideal superconductor does not change at the long-rang phase coherence appearance in $H_{{\rm c}{\rm 2}}$. But it takes place in real (with disorders) superconductors in the consequence of the vortex pinning. The resistivity changes from $\rho _{{\rm n}}$ down to zero at enough low measuring current (at $j < j_{{\rm c}}$) at $H_{{\rm c}{\rm 2}}$ according to the mean field approximation.

The fluctuations change qualitatively the resistivity dependence. The fluctuation decreases the vortex flow resistivity as well as the resistivity above the transition into the Abrikosov state [16]. The fluctuation value increases near the transition. Therefore, a feature ought be expected at the transition from the paraconducting regime to the vortex flow regime. Because the appearance of phase coherence is not connected with the change of the $n_{{\rm s}}$ value, a position of this feature (and the pinning appearance) differs from the $H_{{\rm c}{\rm 2}}$.

\bigskip

\textbf{4. Results of Experimental Investigation of Transition into the Abrikosov State}

\bigskip

According to the fluctuation theory the sharp changes of the resistive properties should be observed at the transition into the Abrikosov state of an ideal superconductor. But the sharp change of the resistive properties is observed in few bulk samples with weak disorder [3,4] only. No sharp change is observed both in thin films with weak disorder [21] and in all samples with strong disorder. The transition into the Abrikosov state becomes smooth with increasing of disorder amount [23]. The absence of the sharp change of the resistive properties means that the transition into the Abrikosov state in thin films and in superconductors with strong disorder is absent or differs from the ideal case.

\begin{figure}
\centering
\includegraphics[width=4.55in,height=2.51in]{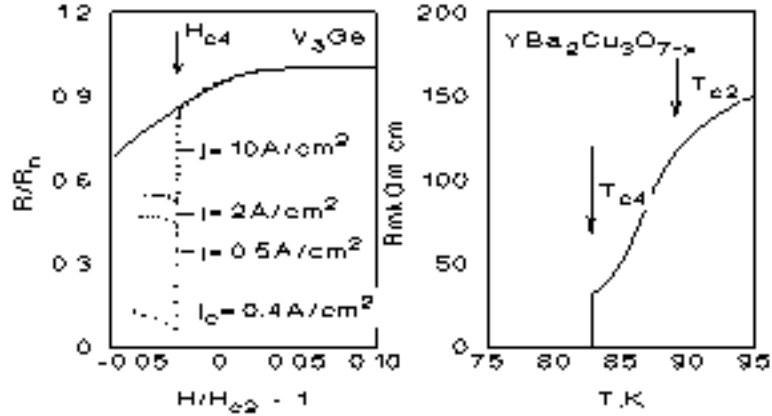}
\caption{ Sharp change of the resistive properties at the transition into the Abrikosov state of $V_{{\rm 3}}Ge$ and YBa$_{{\rm 2}}$Cu$_{{\rm 3}}$O$_{{\rm 7}{\rm -} {\rm x}}$}
\end{figure}

\

4.1. TRANSITION INTO THE ABRIKOSOV STATE IN BULK SUPERCONDUCTORS WITH WEAK DISORDERS. 

\bigskip

The sharp change of the resistive properties was observed first in the paper [3]. This sharp change, observed at $H_{{\rm c}{\rm 4}} < H_{{\rm c}{\rm 2}}$, was interpreted in [3] as the transition into the Abrikosov state. The result of [3] means that this transition in bulk superconductors with weak disorder can be close to the one in a ideal superconductor. The length of the phase coherence changes abruptly from $(\Phi _{{\rm 0}}/H)^{{\rm 1}{\rm /} {\rm 2}} = 10^{{\rm -} {\rm 5}} - 10^{{\rm -} {\rm 6}}$ cm to a sample size $L = 0.01 - 1$ cm in real cases.

The appearance of the vortex pinning (the current-voltage characteristics become non-Ohmic) [3] and the sharp decrease of the vortex flow resistivity [3,20] (Fig.2) are observed in enough homogeneous bulk superconductors at $H_{{\rm c}{\rm 4}}$. Below the transition the resistivity can be equal zero at $j < j_{{\rm c}}$ and the $\rho _{{\rm f}}(H)$ dependence has a minimum [20] or a step (see Fig.3). Such $\rho _{{\rm f}}(H)$ dependence differs qualitatively from the one (\ref{eq4}) predicted by the mean-field approximation [22].

Some authors [24] state that the feature of the $\rho _{{\rm f}}(H)$ dependencies coincide with the peak effect in the critical current. But this is not right. Our investigations [25] have shown that this feature is observed in all enough homogeneous bulk samples both with and without the peak effect (see Fig.3). And only in no enough homogeneous samples the "classical" flux flow resistivity dependence (see [8]) is observed. Therefore this feature ought be considered as universal for homogeneous bulk superconductors. It is explained in [20] by fluctuation influence.

The sharp change of the resistive properties was observed in YBa$_{{\rm 2}}$Cu$_{{\rm 3}}$O$_{{\rm 7}{\rm -} {\rm x}}$ also [4] (see.Fig2). The difference of the $H_{{\rm c}{\rm 2}} -H_{{\rm c}{\rm 4}}$ values in these superconductors corresponds to the LLL scaling law [2]. According to the LLL scaling law [2]: $H_{{\rm c}{\rm 2}} -H_{{\rm c}{\rm 4}} = \varepsilon _{{\rm c}{\rm 4}}H_{{\rm c}{\rm 2}}(0)Gi^{{\rm 1}{\rm /} {\rm 3}{\rm }}(H_{{\rm c}{\rm 2}}/H_{{\rm c}{\rm 2}}(0))^{{\rm 2}{\rm /} {\rm 3}} (T/T_{{\rm c}})^{{\rm 2}{\rm /} {\rm 3}}$. The experimental value $\varepsilon _{{\rm c}{\rm 4}} \quad  \approx  1$ for superconductors with different $Gi$ and $H_{{\rm c}{\rm 2}}(0)$ values: $Nb_{{\rm 9}{\rm 4}{\rm .}{\rm 3}}Mo_{{\rm 5}{\rm .}{\rm 7}}$, $Gi = 10^{{\rm -} {\rm 9}}$, $H_{{\rm c}{\rm 2}}$(0) = 0.8 T; $V_{{\rm 3}}Ge$, $Gi = 10^{{\rm -} {\rm 6}}$, $H_{{\rm c}{\rm 2}}(0)$ = 12 T; YBa$_{{\rm 2}}$Cu$_{{\rm 3}}$O$_{{\rm 7}{\rm -} {\rm x}}$, $Gi = 10^{{\rm -} {\rm 2}}$, $H_{{\rm c}{\rm 2}}$(0) = 200 T. Consequently, the transition from the mixed state without the phase coherence into the Abrikosov state is observed both in [3] and in [4]. Result of [26] shows that this transition in bulk superconductors with weak disorder can be first order indeed.

\begin{figure}
\centering
\includegraphics[width=4.14in,height=2.80in]{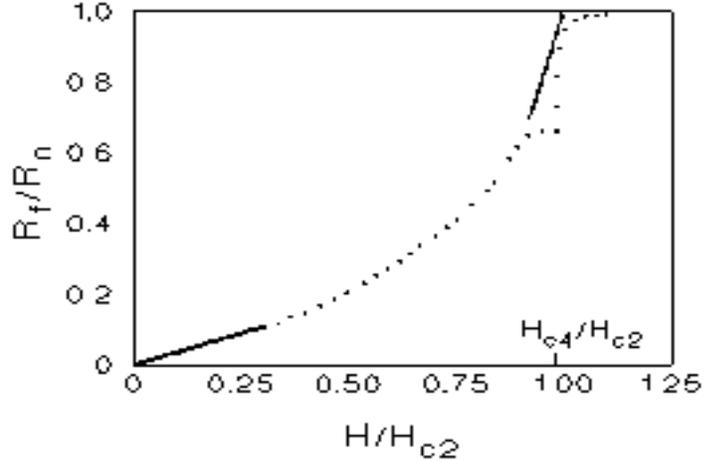}
\caption{ Magnetic dependence of the vortex flow resistivity of $Ti_{{\rm 8}{\rm 4}}Mo_{{\rm 1}{\rm 6}}$ alloy. The peak effect in the critical current is absent in this sample. Lines are mean field approximation theoretical dependencies.} 
\end{figure}

\bigskip

4.2. ABSENCE OF THE TRANSITION INTO THE ABRIKOSOV STATE IN TWO-DIMENSIONAL SUPERCONDUCTORS WITH WEAK DISORDER DOWN TO VERY LOW MAGNETIC FIELD. 

\bigskip

The absence of any features of the resistive dependence of amorphous $Nb_{{\rm 1}{\rm -} {\rm x}}O_{{\rm x}}$ films was interpreted in our paper [21] as the absence of the transition into the Abrikosov state down to very low magnetic fields. The vortex pinning appearance can not be observed if the measuring current is extremely high in comparison to the critical current. But the feature of the vortex flow resistivity can be observed at a high measuring current. The absence of the long-rang phase coherence below $H_{{\rm c}{\rm 2}}$ in these films is confirmed by a investigation of the nonlocal resistivity [27].

The feature of the vortex flow resistivity are observed in the $a-NbGe$ films with an intermediate strength of disorder [24]. This feature are observed below the field value where the vortex pinning appears, as well as in bulk superconductors [20]. Consequently, the phase coherence appearance occurs in films with an intermediate strength of disorder. Comparison of the results of [21] and [24] shows that the phase coherence appearance in thin film is not universal, but that it depends on the amount of disorder. The dependence of the position of the transition into the Abrikosov state of two-dimensional superconductors on the amount of disorder is predicted by results of the work [28].

\bigskip

4.3 PHASE COHERENCE APPEARANCE IN SUPERCONDUCTORS WITH STRONG DISORDERS. RETURN TO THE MENDELSSOHN MODEL

\bigskip

It is obvious that the long-rang phase coherence is appeared in superconductors with strong and intermediate strength of disorder, because the pinning effect is observed in these cases. But the phase coherence appearance in these cases differs from the ideal case. The length of the phase coherence does not change by jump but increases gradually with the magnetic field (or the temperature) decreasing.

The smooth transition observed in most samples can be connected with an increase of the effective dimensionality of the fluctuation (from zero to one in film) in samples with strong disorder. The limit case of strong disorder is the Mendelssohn's "sponge" (see [29]). The Mendelssohn's "sponge" is a model proposed in 1935 year for explanation of the magnetic properties of some superconductors (now we call these superconductors as type II superconductors). Mendelssohn has assumed that the magnetic field can penetrate inside superconducting region because this region is superconducting sponge.

The Abrikosov state in thin film is similar to the two-dimensional Mendelssohn sponge. The main difference is that the Abrikosov vortex destroys superconductivity near itself (in a region with $r < \xi $) in the Abrikosov model whereas in the Mendelssohn model it occupies a nonsuperconducting region. We may consider real superconductors with disorder as intermediate cases between the Mendelssohn's and Abrikosov's models. A thin film with strong disorder can be considered as a Mendelssohn sponge with a variable width of superconducting threads [30]. 

The length of phase coherence increases smoothly with temperature decreasing in a Mendelssohn sponge as well as a one-dimensional superconductor [31]. In consequence of this the resistive dependence is smooth also both in a Mendelssohn sponge and in a one-dimensional superconductor.

The length on the phase coherence in superconductor with strong disorders 
can begin to increase already above $H_{{\rm c}{\rm 2}}$ [30]. At a value of 
the magnetic field $< H_{{\rm c}{\rm 2}}$ a crossover to the vortex creep 
regime takes place. This crossover can be interpreted as a consequence of 
the increasing of the phase coherence length up to sample size. 

\bigskip

\textbf{5.Theories of the Vortex Lattice Melting and Theories of the Vortex Liquid Solidification}

\bigskip

It is proposed in vortex lattice melting theories (see [1,32]) that the Abrikosov state is the vortex lattice. This opinion is based on the Abrikosov solution [12]. But according to the Maki-Takayama result [16] this solution is not valid in thermodynamic limit. Consequently, the vortex lattice melting theories are unsatisfactory in principle, since they start from the state in which the translation symmetry has been broken by hand [33]. Therefore, some theorists consider no the vortex lattice melting but the solidification transition of vortexes (see for example [34,35]). They do not propose the Abrikosov state existence, but try to find the transition to it. 

The phase coherence is defined by the correlation function in the solidification theory. This definition is logical contradictory because according to this definition the long-rang phase coherence can not exist without the crystalline long-rang order of the vortex lattice. It is claimed in some works [36] that the phase coherence can remain short-ranged even in the vortex solid phase. Therefore the solidification theories try to find the transition into the Abrikosov state as the vortex lattice but no as the mixed state with long-rang phase coherence. Most authors find this transition (see for example [34]). And few authors (see for example [35]) only state that this transition is absent. I agree with few authors [35]. But the absence of the solidification transition does not mean that the transition into the Abrikosov state is absent, because the definition of the phase coherence by the correlation function is unsuited for the mixed state. According to the right definition of the phase coherence the Abrikosov state is the mixed state with long-rang phase coherence but no the vortex lattice.

\bigskip

\textbf{6. Conclusions}

\bigskip

Because the appearance of the vortex pinning and the feature of the vortex flow resistivity should be observed at the appearance of long-rang phase coherence it is obvious that the interpretation proposed in [3] is right. The sharp change of the resistive observed in bulk superconductors with weak disorders at a field $H_{{\rm c}{\rm 4}} < H_{{\rm c}{\rm 2}}$ is the transition from the mixed state without phase coherence into the Abrikosov state.

But this transition observed in YBa$_{{\rm 2}}$Cu$_{{\rm 3}}$O$_{{\rm 7}{\rm -} {\rm x}}$ was interpreted in [4,26] and other papers as the vortex lattice melting. This interpretation is very popular, but it can not be right. The vortex liquid is the mixed state with long-rang phase coherence, because the existence of the vortexes is evidence of the long-rang phase coherence. Consequently, the first order phase transition from the vortex liquid into the mixed state without the phase coherence must be observed above the vortex lattice melting in superconductors with weak disorders. The sharp change of the resistive properties must be observed at this transition. But no sharp change is observed above $H_{{\rm c}{\rm 4}}$ both in conventional superconductors [3] and in YBa$_{{\rm 2}}$Cu$_{{\rm 3}}$O$_{{\rm 7}{\rm -} {\rm x}}$ [4].

The sharp transition into the Abrikosov state predicted by the fluctuation theory in ideal case is observed in bulk superconductors with weak disorder [3,4] only. No sharp transition is observed in thin films with weak disorder [21]. This difference from bulk superconductor can be explained by difference of the fluctuation value in three- and two-dimensional superconductors [16]. The nature of the step in the magnetization dependence of layered HTSC [37] is some questionable because no sharp transition is observed in thin film of conventional superconductors [21]. This step can not be interpreted as a universal transition in three- or two-dimensional superconductor. It is most probably a consequence of a transition from two- to three dimensional behavior in layered HTSC. Such interpretation was proposed in some papers.

The second critical field $H_{{\rm c}{\rm 2}}$ is not a critical point not only in superconductors with weak disorder but also in superconductors with strong disorder. The phase coherence appears below $H_{{\rm c}{\rm 2}}$ in superconductors with weak disorder [3,4] and above $H_{{\rm c}{\rm 2}}$ in superconductors with strong disorder [30].

The smooth phase coherence appearance in superconductors with strong disorder can be explained qualitatively by the increasing of the effective dimensionality of the fluctuation.

\bigskip

\textbf{Acknowledgement}

\bigskip

I thank the International Association for the Promotion of Co-operation with Scientists from the New Independent States (Project INTAS-96-0452) and the National Scientific Council on "Superconductivity" of SSTP "ADPCM" (Project 98013) for financial support. 

\bigskip

\textbf{References}

\bigskip

1. Brandt, E.H. (1995) The flux-line lattice in superconductors, \textit{Rep.Progr.Phys.} \textbf{58}, 1465-1594.

2. Nikulov, A.V. (1997) Fluctuation effects in mixed state of type II superconductors, in M.Aussloos and A.A.Varlamov (eds.), \textit{Fluctuation Phenomena in High Temperature Superconductors,} Kluwer Academic 
Publishers, Dordrecht, pp.271-277. 

3. Marchenko, V.A. and Nikulov, A.V. (1981) Magnetic field dependence of the 
electrical conductivity in V$_{{\rm 3}}$Ge in the vicinity of H$_{{\rm c}{\rm 2}}$, \textit{Pisma Zh.Eksp.Teor.Fiz.} \textbf{34}, 19-21 (\textit{JETP Lett.} \textbf{34}, 17-19).

4. Safar, H., Gammel, P.L., Huse, D.A., Bishop, D.J., Rice, J.P., and 
Ginzberg, D.M. (1992) Experimental evidence for a first-order 
vortex-lattice-melting transition in untwinned single crystal YBa$_{{\rm 2}}$Cu$_{{\rm 3}}$O$_{{\rm 7}}$, \textit{Phys.Rev.Lett.} \textbf{69,} 824-827.

5. Nikulov, A.V., (1990) On phase transition of type-II superconductors into the mixed state, \textit{Supercond.Sci.Technol.} \textbf{3}, 377-380. 

6. Bishop, D. (1996) Has the fat lady sung? \textit{Nature} \textbf{382}, 760-761.

7. Nelson, D.R. (1995) Vortex lattice melts like ice, \textit{Nature} \textbf{375}, 356-357.

8. Huebener, R.P. (1979) \textit{Magnetic Flux Structures in Superconductors,} Springer-Verlag, Berlin.

9. Tinkham, M. (1975) \textit{Introduction to Superconductivity}, McGraw-Hill Book Company, New-York.

10. Chen, D.-X. et al., (1998) Nature of the driving force on an Abrikosov 
vortex, \textit{Phys.Rev. B} \textbf{57}, 5059-5062.

11. Kleiner, W.H., Roth, L.M., and Autler S.H. (1964) Bulk solution of 
Ginzburg-Landaw equations for type II superconductors: Upper critical field 
region, \textit{Phys.Rev. A} \textbf{133}, 1226-1227.

12. Abrikosov, A.A. (1957) On the magnetic properties of superconductors of the second group, \textit{Zh.Eksp.Teor.Fiz.} \textbf{32}, 1442-1452 (\textit{Sov.Phys.-JETP} \textbf{5}, 1174-1184). 

13. De Gennes, P.G. (1966) \textit{Superconductivity of Metals and Alloys}, Pergamon Press.

14. Lee, P.A. and Shenoy, S.R. (1972) Effective dimensionality change of fluctuation in superconductors in a magnetic field, \textit{Phys.Rev.Lett}. \textbf{28}, 
1025-1028.

15. Eilenberger, G. (1967) Thermodynamic fluctuations of the order parameter in type II superconductors near the upper critical field H$_{{\rm c}{\rm 
2}}$, \textit{Phys.Rev}. \textbf{164}, 628-635.

16. Maki, K. and Takayama,H. (1971) Thermodynamic fluctuation of the order parameter in the vortex state of type II superconductors, \textit{Prog.Theor.Phys.} \textbf{46}, 
1651-1665.

17. Essmann, U. and Trauble, H. (1967) Direct observation of a triangular flux-line lattice in type II superconductors by a Bitter method, \textit{Phys.Lett. A} 
\textbf{24,} 526-532.

18. Larkin, A.I. (1970) Effect of inhomogeneities on the structure of the mixed state of superconductors, \textit{Zh.Eksp.Teor.Fiz.} \textbf{58}, 1466-1470 (\textit{Sov.Phys.-JETP} \textbf{31}, 784-788).

19. Maki, K. and Thompson, R.S. (1989) Fluctuation contribution to flux-flow conductivity, \textit{Physica} C \textbf{162-164}, 275-279.

20. Marchenko, V.A. and Nikulov, A.V. (1981) Fluctuation conductivity in V$_{{\rm 3}}$Ge near the second critical field, \textit{Zh.Eksp.Teor.Fiz.} \textbf{80}, 745-750 (\textit{Sov.Phys.-JETP} \textbf{53}, 377-381). 

21. Nikulov, A.V., Remisov, D.Yu., and Oboznov, V.A. (1995) Absence of the 
transition into Abrikosov vortex state of two-dimensional type-II 
supercondutor with weak pinning, \textit{Phys.Rev.Lett.} 75, 2586-2589.

22. Gor'kov, L.P. and Kopnin, N.B. (1975) Flux flow and electrical 
resistivity of type II superconductors in a magnetic field, \textit{Usp.Fiz.Nauk} \textbf{116}, 
413-448 (\textit{Sov. Phys.Usp.} \textbf{18}, 496-531.)

23. Fendrich, J.A. et al., (1995) Vortex liquid state in an electron irradiated untwinned YBa$_{{\rm 2}}$Cu$_{{\rm 3}}$O$_{{\rm 7}}$ crystal, 
\textit{Phys.Rev.Lett}. \textbf{74}, 1210-1213.

24. Theunissen, M.H. and Kes, P.H. (1997) Resistive transitiond of thin film superconductors in a magnetic field, \textit{Phys.Rev.B} \textbf{55}, 15183-15190.

25. Nikulov, A.V. (1985) Fluctuation phenomena in bulk type-II superconductors near second critical field, \textit{Thesis} (Inst. of Solid State Physics, USSR Academy of Sciences, Chernogolovka).

26. Schilling, A. et al., (1996) Calorimetric measurement of the latent heat of vortex-lattice melting in untwinned YBa$_{{\rm 2}}$Cu$_{{\rm 3}}$O$_{{\rm 7}}$, \textit{Nature} \textbf{382}, 791-793.

27. Nikulov, A.V., Dubonos, S.V. and Koval,Y.I. (1997) Destruction of the phase coherence by magnetic field in the fluctuation region of thin superconducting films, \textit{J.Low Temp.Phys.} \textbf{109}, 643-652.

28. Nikulov, A.V. (1995) Existence of Abrikosov vortex state in two-dimensional type-II superconductors without pinning, \textit{Phys.Rev. B} \textbf{52}, 10429-10432.

29. Shoenberg, D. (1952) \textit{Superconductivity}, Cambrige.

30. Nikulov, A.V. Phase coherence appearence in thin superconducting film with strong disorders. The return to the Mendelssohn model.  \ http://publish.aps.org/eprint/gateway/epget/aps1998mar20\_002

31. Grunberg, L.W. and Gunther, L. (1972) Critical behavior of one-dimensional superconductors: an exact solution, \textit{Phys.Lett}. A \textbf{38}, 463-464.

32. Blatter, G., Feigel'man, M.V., Geshkenbein, V.B., Larkin, A.I., and Vinokur V.M. (1994) Vortex in High-T$_{{\rm c}}$ Superconductors, \textit{Rev.Mod.Phys.} \textbf{66}, 1125-1388.

33. Tesanovic,Z. (1991) Nature of the superconducting transition in the presence of a magnetic field, \textit{Phys.Rev. B} \textbf{44}, 12635-12638.

34. Hu,J. and MacDonald,A.H. (1995) Participation-ratio entropy and critical fluctuations in the thermodynamics of pancake vortices, \textit{Phys.Rev}. B \textbf{52}, 
1286-1289.

35. Lee,H.H. and Moore,M.A. (1994) Monte Carlo studies of the two-dimensional vortex liquid: absence of transition and dynamical 
properties, \textit{Phys. Rev}. B \textbf{49}, 9240-9243.

36. Ikeda,R. (1996) Superconducting ordering in vortex states of clean type II superconductors with pinnings, \textit{J. Phys.Soc.Jpn.} \textbf{65}, 3998-4017.

37. Zeldov, E. et al., (1995) Thermodynamic observation of first-order vortex- lattice melting transition in Bi$_{{\rm 2}}$Sr$_{{\rm 
2}}$CaCu$_{{\rm 2}}$O$_{{\rm 8}}$, \textit{Nature} \textbf{375}, 373-376.

\end{document}